\documentclass{article}
\usepackage{mrs2005,epsfig}
\begin{document} 
\title{THE SOURCES OF THE COSMIC INFRARED BACKGROUND}

\author{Guilaine Lagache, Herv\'e Dole, Jean-Loup Puget}
\affil{Institut d'Astrophysique Spatiale (IAS), B\^atiment 121, F-91405 Orsay (France); Universit\'e Paris-Sud 11 and CNRS (UMR 8617)
}

\begin{abstract} 
The discovery of the Cosmic Infrared Background (CIB) in 1996,
together with recent cosmological surveys
from the mid-infrared to the millimeter have revolutionized
our view of star formation at high redshifts.
It has become clear, in the last decade, that a population of galaxies 
that radiate most of their power in the far-infrared (the 
so-called ``infrared galaxies'') contributes an important part
of the whole galaxy build-up in the Universe. Since 1996, detailed (and often painful)
investigations of the high-redshift infrared galaxies have resulted in the
spectacular progress reviewed in this paper. Among others, we emphasize a new 
{\it Spitzer} result based on a Far-IR stacking analysis of mid-IR sources. 
\end{abstract} 
 
\section{Introduction} 
\begin{figure}[here]  
\vspace*{1.25cm}  
\begin{center}
\epsfig{figure=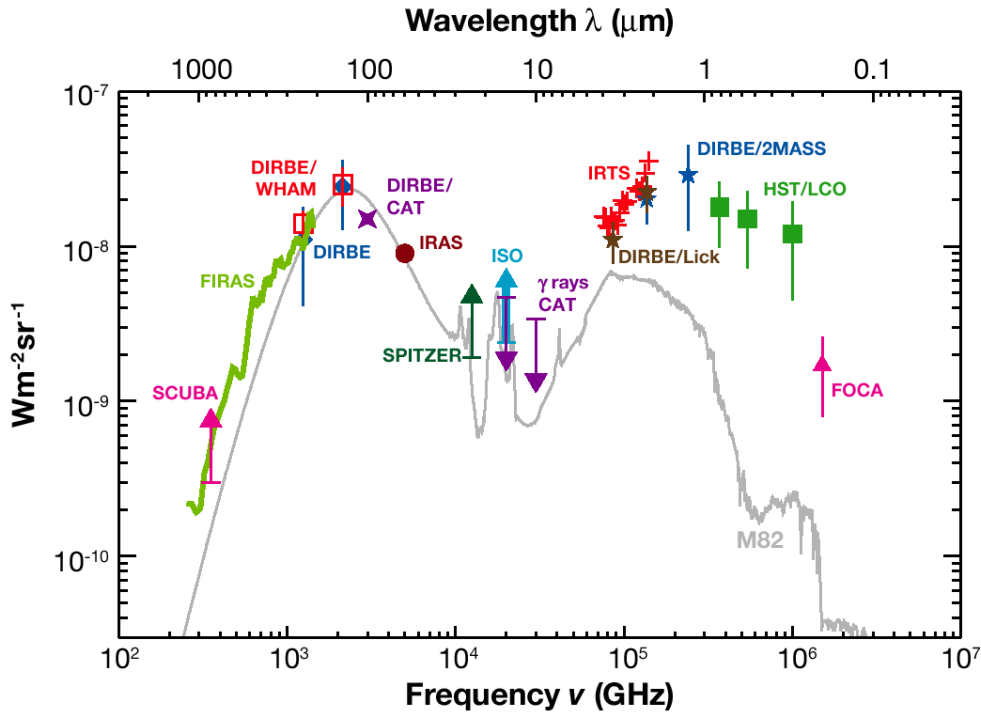,width=13cm}  
\end{center}
\vspace*{0.25cm}  
\caption{\label{CIB_tot} The extragalactic background over three decades in frequency from the
near UV to millimeter wavelengths. Only strongly constraining
measurements have been reported. We show for comparison in grey an SED
of M82 (Chanial, 2003)-- a starburst galaxy at L=3$\times$ 10$^{10}$
L$_{\odot}$ -- normalised to the peak of the CIB at 140~$\mu m$. 
References for data points are given in Lagache et al. (2005). 
} 
\end{figure}

 The CIB can be defined as the part of the present radiation content of the Univers made essentially of the 
long-wavelength output from all sources throughout the history of the Universe.
Fig. \ref{CIB_tot} shows the cosmic background due to sources from the UV up to the mm. It has two maxima, one in the 
optical and one in the far-IR with roughly equal brightness and with a minimum at 5 microns. This minimum 
is created by the decrease of brightness of the stellar component with $\lambda$ combined with the 
rising brightness of the dust. The CIB is defined as the cosmic background at $\lambda$ longward of this 
minimum. An understanding of the nature and resdshift distribution of the sources of the CIB, 
although relatively new, is an integral part of our understanding of the formation and evolution of galaxies.\\

%
%

The {\it IRAS} satellite, launched in 1983 gave for the first time
a proper census of the infrared emission of galaxies at low redshift.
The Luminosity Function (LF) at 60 and 100 $\rm \mu m$ is
dominated by $L_\star$ spiral galaxies as could be expected 
-- the reradiated stellar luminosity absorbed by dust.
In addition, a high-luminosity tail of luminous galaxies was found
(e.g. Sanders \& Mirabel 1996). This high-luminosity tail can be approximated by a 
power-law, $\Phi (L) \propto L_{\rm IR}^{2.35}$, which gives a space density for the most
luminous infrared sources well in excess of predictions based on the optical
LF. These sources comprise the Luminous Infrared Galaxies, LIRGs, and the
ULIRGs with luminosities 11$<$log(L$_{\rm IR}$/L$_{\odot}$)$<$12
and log(L$_{\rm IR}$/L$_{\odot}$)$>$12, respectively. These galaxies are often associated with interacting
or merging, gas-rich disks.
They do not dominate the energy production locally.
As an example, the total infrared luminosity from these
galaxies in the {\it IRAS} Bright Galaxy Sample accounts for only 
$\sim$6\% of the infrared emission in the local Universe
(Soifer \& Neugeubauer, 1991). The situation
changes dramatically at higher redshift where
these galaxies fully dominate the infrared energy output.
We just give here for illustration a very simple argument.
We  know that locally, the infrared output of galaxies is only one third of the
optical output. In contrast, we observe in the cosmic background
a power in the infrared comparable to the power in the optical.
This implies that infrared galaxies grow more luminous
with increasing $z$ faster than do optical galaxies.\\

In addition to the CIB, the observations relevant to the problem of star and galaxy formation at high $z$ 
from the mid-IR ($\sim$10 $\mu$m) to the mm are the following:
\par\medskip\noindent
$\bullet$ Deep number counts obtained at 15, 24, 90, 160, 170, 350/450, 850, 1200 $\mu$m
(e.g. Dole et al. 2001,  Elbaz et al. 2002, Scott et al. 2002, 
Dole et al. 2004, Greve et al. 2004, H\'eraudeau et al. 2004, Papovich et al. 2004,).
\par\medskip\noindent
$\bullet$ Luminosity function up to $z \sim$1 (e.g. Le Floc'h et al. 2005, P\'erez-Gonz\'alez et al. 2005)
\par\medskip\noindent
$\bullet$ Clustering properties (e.g. Blain et al. 2004 for a first attempt at 850 $\mu$m)
\par\medskip\noindent
$\bullet$ Power spectra of the unresolved background 
(Lagache \& Puget 2000, Matsuhara et al. 2000, Miville-Desch\^enes et al. 2002, Maloney et al.
2005)
\par\medskip\noindent
$\bullet$ Identifications and multi-wavelength observations of IR galaxies
\par\medskip\noindent
We outline here the main results in this field. More details are given in 
Lagache et al. (2005).

\section{\label{zdistribCIB} Redshift distribution of the sources of the CIB}

\begin{figure}[here]  
\vspace*{1.25cm}  
\begin{center}
\epsfig{figure=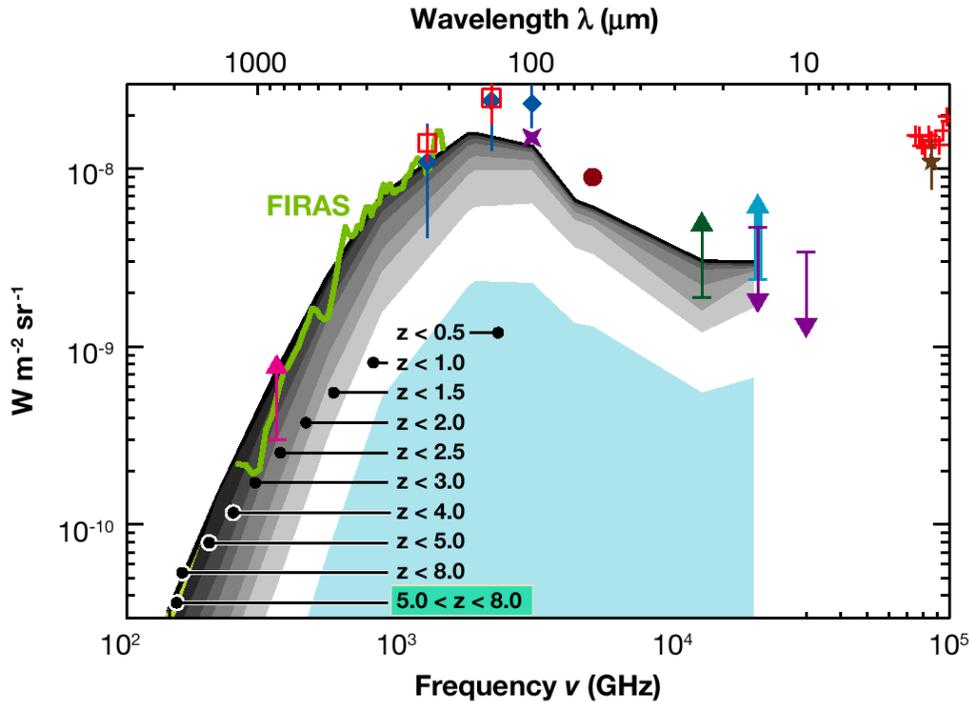,width=13cm}  
\end{center}
\vspace*{0.25cm}  
\caption{\label{fig:CIB_fraction_spec} Cumulative contribution to the CIB of galaxies at various
redshifts from 0.5 to 8, from the model of Lagache
et al. (2004). Measurements of the CIB are reported with the same
symbols as in Figure~\ref{CIB_tot}. 
} 
\end{figure}

An other important property to note from the observation of the CIB is that the
slope of the long wavelength part of the CIB, $I_\nu \propto
\nu^{1.4}$ (Gispert et al. 2000), is much less steep than the long
wavelengths spectrum of galaxies (as illustrated in Figure
\ref{CIB_tot} with the M82 SED).  This implies that the millimeter
CIB is not due to the millimeter emission of the galaxies that account
for the peak of the CIB ($\simeq 150 \mu$m). In  fact, contributions from galaxies
at various redshifts are needed to fill the CIB SED shape.  The bulk
of the CIB in energy, i.e., the peak at about 150 $\rm \mu$m, is not
resolved in individual sources (as discussed in Sect. \ref{Resolved_CIB}) 
but one dominant contribution at the
CIB peak can be inferred from the ISOCAM deep surveys. ISOCAM galaxies
with a median redshift of $\sim$0.7 resolve about 80\% of the CIB at
15~$\rm \mu$m. Elbaz et al. (2002) separate the 15 $\rm \mu$m galaxies
into different classes (ULIRGs, LIRGs, Starbursts, normal galaxies and
AGNs) and extrapolate the 15 $\rm \mu$m fluxes to 140~$\rm \mu$m using
template SEDs. A total brightness of (16$\pm$5) nW m$^{-2}$ sr$^{-1}$
is found, which makes up about two thirds of the CIB observed at 140
$\rm \mu$m by COBE/DIRBE. Hence, the galaxies detected by ISOCAM are
responsible for a large fraction of energy of the CIB. 
However, these galaxies make little contribution to the
CIB in the millimeter and submillimeter. There, the CIB must be
dominated by galaxies at rather high redshift for which the SED peak
has been shifted. 
The redshift contribution to the CIB is illustrated
in Figure \ref{fig:CIB_fraction_spec}.  We clearly see that the
submillimeter/millimeter CIB contains information on the total energy
output by the high-redshift galaxies ($z>$2). This is supported by the
redshift distribution of the SCUBA sources at 850 $\rm \mu$m with
S$_{850}$$\ge$3 mJy that make about 30\% of the CIB and have a median
redshift of 2.2 (Chapman et al. 2005).

\section{\label{Resolved_CIB} The Status of Deep Surveys: Resolved Fraction of the $>$10 $\rm \mu$m CIB}

Many surveys from the mid-infrared to the millimeter have aimed to
resolve the CIB into discrete sources (Table \ref{tab:reolved_CIB}).
From Table \ref{tab:reolved_CIB}, we see that the most constraining surveys
in term of resolving the CIB are those at 15, 24 and 850 $\rm \mu$m.
Moreover, the capabilities of these surveys to find high-$z$ objects
are the best among all other existing surveys (see Lagache et al. 2005). 
These surveys probe the CIB in well-defined and distinct redshift ranges, 
with median redshifts of 0.7 (Liang et al. 2004), $\sim$1 
(Caputi et al. 2005 and L. Yan,
private communication), and 2.2 (Chapman et al. 2005) at 15, 24 and
850~$\rm \mu$m, respectively. Such well-defined redshift ranges are due 
to the peculiar shape of K-corrections at these wavelengths.

\section{Detailed properties of the sources of the CIB}

\subsection{The $0.5<z<1.5$ galaxies}
At the time of writing, most of the detailed informations
on dusty galaxies in the $0.5 \le z \le 1.5$ redshift range comes from
galaxies selected with the ISOCAM cosmological surveys 
at 15 $\rm \mu$m and the multi-wavelength analysis of detected sources. 
Shallow, deep and ultra-deep surveys were performed in various
fields including the Lockman hole, Marano, northern and southern
Hubble Deep Field (HDF), and Canada-France Redshift Survey (CFRS)
(e.g., Aussel et al. 1999; Flores et al. 1999; Lari et al. 2001;
Gruppioni et al. 2002; Mann et al. 2002; Elbaz \& Cesarsky 2003; Sato
et al. 2003). Deeper images have been made in the direction of distant
clusters (e.g., Metcalfe et al. 2003).  Finally, the bright end of the
luminosity function was explored by the ELAIS survey (e.g., Oliver et
al. 2000).  The deepest surveys reach a completeness limit of about
100 $\rm \mu$Jy at 15 $\rm \mu$m (without lensing). The most relevant
data to this section are the deep and ultra-deep surveys.\\

To find out the nature and redshift distribution of the 15 $\rm \mu$m
deep survey sources, many followup observations have been conducted
including HST imaging and VLT spectroscopy.  With a point-spread
function full width at half of maximum of 4.6 arcsec at 15~$\rm \mu$m,
optical counterparts are easily identified. Redshifts are found using
emission and/or absorption lines. From field to field, the median
redshift varies from 0.52 to 0.8, a quite large variation due to
sample variance. Most of ISOCAM galaxies have redshifts between $\sim$ 0.3 and
1.2, consistent with Figure \ref{fig:CIB_fraction_spec}. About
85$\%$ of the ISO galaxies show obvious strong emission lines (e.g.,
[OII] 3723, H$_{\gamma}$, H$_{\beta}$, [OIII] 4959, 5007). These lines
can be used as a diagnostic of the source of ionization and to
distinguish the HII-region like objects from the Seyferts and LINERs.
Most of the objects are found to be consistent with HII regions, e.g.,
from Liang et al. (2004) and exhibit low ionization level
([OIII]/H$_{\beta} <$3). From emission lines studies, the AGN fraction
is quite low, $\sim$20 $\%$. This is consistent with X-ray
observations of ISOCAM sources showing that AGNs contribute at most
17$\pm$6$\%$ of the total mid-infrared flux (Fadda et
al. 2002). Assuming template SEDs typical of star-forming and
starburst galaxies, 15 $\rm \mu$m fluxes can be converted into total
infrared luminosities, L$_{\rm IR}$ (between 8 and 1000 $\rm
\mu$m). About 75$\%$ of the galaxies dominated by the star formation
are either LIRGs or ULIRGs. The remaining 25\% are nearly equally
distributed among either "starbursts" (10$^{10}<$ L$_{\rm IR}<$
10$^{11}$ L$_{\odot}$) or "normal" (L$_{\rm IR}<$ 10$^{10}$
L$_{\odot}$) galaxies. The median luminosity is about
3$\times$10$^{11}$ L$_{\odot}$. ULIRGs and LIRGs contribute to about
17$\%$ and 44$\%$ to the CIB at 15 $\rm \mu$m, respectively (Elbaz et
al. 2002). This suggests that the star formation density at $z<$1 is
dominated by the abundant population of LIRGs (see also Le Floc'h et
al. 2005).  
Because of large extinction in LIRGs and ULIRGs, the infrared data
provide more robust SFR estimate than UV tracers. The extinction
factor in LIRGs averages to A$_V$ $\sim$2.8 at $z\sim$0.7 (Flores et
al. 2004). It is much higher than that of the local star-forming
galaxies for which the median is 0.86 (Kennicutt 1992). Assuming
continuous burst of age 10-100 Myr, solar abundance, and a Salpeter
initial mass function, the SFR can be derived from the infrared
luminosities (Kennicutt 1998).
Typical LIRGs form stars at $\ge$20 M$_{\odot}$ year$^{-1}$.  The
median SFR for the 15 $\rm \mu$m galaxies is about 50 M$_{\odot}$
year$^{-1}$, a substantial factor larger than that found for
faint-optically selected galaxies in the same redshift range. \\

\begin{table}[top]
\begin{center}
\begin{tabular}{|l|c|l|} \hline
Survey  & Fraction of resolved CIB & Reference \\ \hline
ISOCAM 15$\mu$m & 80\% & Elbaz et al. 2002 \\
SPITZER 24$\mu$m & 70\% & Papovich et al. 2004 \\
SPITZER 70$\mu$m & 23\% & Dole et al. 2004 \\
ISOPHOT 90$\mu$m & $<$5\% & H\'eraudeau et al. 2004 \\
SPITZER 160$\mu$m & 7\% & Dole et al. 2004 \\
ISOPHOT 170$\mu$m & $<$5\% & Dole et al. 2001 \\
SCUBA 450$\mu$m & 15\% & Smail et al. 2002 \\
SCUBA 850$\mu$m & 60\% (S$>$1 mJy) & Smail et al. 2002 \\
                & 30\% (S$>$3 mJy) & Lagache et al. 2005 \\
MAMBO 1.2 mm & 10\% &  Greve et al. 2004 \\ \hline
\end{tabular}
\end{center}
\caption{\label{tab:reolved_CIB} CIB resolved fraction}
\end{table}

The other fundamental parameter characterizing the sources of the peak
of the infrared background is their stellar mass content that traces
the integral of the past star formation activity in the galaxies and
is a natural complement to the instantaneous rate of star
formation. The stellar masses can be obtained using spectral synthesis
code modeling of the UV-optical-near infrared data or, more simply
using the mass-to-luminosity ratio in the K-band. The derived stellar
masses for the bulk of ISOCAM galaxies range from about 10$^{10}$ to
3$\times$10$^{11} $M$_{\odot}$, compared to 1.8$\times$10$^{11}$
M$_{\odot}$ for the Milky Way. As expected from the selection based on
the LW3 flux limit -- and thus on the SFR -- masses do not show
significant correlation with redshift (Franceschini et al. 2003).  An
estimate of the time spent in the starburst state can be obtained by
comparing the rate of ongoing star formation (SFR) with the total mass
of already formed stars: $t[years]=M/$SFR.  Assuming a constant SFR,
$t$ is the timescale to double the stellar mass.  For LIRGs at
$z>$0.4, $t$ ranges from 0.1 to 1.1 Gy with a median of about 0.8 Gyr
(Franceschini et al. 2003; Hammer et al. 2005). From $z=1$ to $z=0.4$
(i.e., 3.3 Gyr), this newly formed stellar mass in LIRGs corresponds
to about 60$\%$ of the $z=1$ total mass of intermediate mass
galaxies. The LIRGs are shown to actively build up their metal
content. In a detailed study, Liang et al. (2004) show that, on
average, the metal abundance of LIRGs is less than half of the $z \sim
0$ disks with comparable brightness. Expressed differently, at a given
metal abundance, all distant LIRGs show much larger B luminosities
than local disks. Assuming that LIRGs eventually evolve into the local
massive disk galaxies, Liang et al. (2004) suggest that LIRGs form
nearly half of their metals and stars since $z\sim$1.\\

Finally, morphological classification of distant LIRGs is essential to
understand their formation and evolution. Zheng et al. (2004)
performed a detailed analysis of morphology, photometry, and color
distribution of 36 (0.4$<z<$1.2) ISOCAM galaxies using HST
images. Thirty-six percents of LIRGs are classified as disk galaxies
with Hubble type from Sab to Sd; 25\% show concentrated light
distributions and are classified as Luminous Compact Galaxies (LCGs);
22\% display complex morphology and clumpy light distributions and are
classified as irregular galaxies; only 17\% are major ongoing mergers
showing multiple components and apparent tidal tails. This is clearly
different from the local optical sample of Nakamura et al. (2004) in
the same mass range in which 27\%, 70\%, $<$2\%, 3\% and $<$2\% are
E/S0, spirals, LCGs, irregulars and major mergers
respectively.  For most compact LIRGs, the color maps reveal
a central region strikingly bluer than the outer regions. These blue
central regions have a similar size to that of bulges and a color
comparable to that of star-forming regions. Because the bulge/central
region in local spiral is relatively red, such a blue core structure
could imply that the galaxy was forming the bulge (consistent with
Hammer et al. 2001). It should be noticed that they find all LIRGs
distributed along a sequence that relates their central color to their
compactness.  This is expected if star formation occurs first in the
center (bulge) and gradually migrate to the outskirts (disk), leading
to redder colors of the central regions as the disk stars were
forming (see also Hammer et al., this conference).

\subsection{The $z>1.5$ galaxies}
Analysis of the CIB in the light of the ISO observations shows that, as we go to wavelengths 
much longer than the emission peak,
the CIB should be dominated by galaxies at higher redshifts.
The main source of observations of the submm/mm CIB are the SCUBA submillimeter
observations at 850~$\mu$m and 450~$\mu$m (see Blain et al 2002 for a
review) and observations from the MAMBO instrument on the IRAM 30-m
telescope at 1.2 mm (Greve et al. 2004).
The first obvious question when investigating the nature of the
submillimeter galaxies (SMGs) is thus their redshift distribution. The
rather low angular resolution of the submillimeter and millimeter
observations made identifications with distant optical galaxies an
almost impossible task without an intermediate identification. This is
provided by radio sources observed with the VLA with 10 times better
angular resolution. The tight correlation between far-infrared
luminosity and radio flux (Helou et al. 1985; Condon 1992) provides
the needed link.  This then allows us to get optical identifications
and redshift measurements using 10-m class telescopes.  Confirmation
of these identifications can then be obtained through CO line
observations with the millimeter interferometers such as the Plateau
de Bure interferometer.  The redshift deduced from the optical lines
is confirmed by the CO observations.  So far, only a handful of cases
have gone through this whole chain of observations (e.g., Genzel et
al. 2003; Greve et al. 2005; Neri et al. 2003), but a high success
rate gives confidence in the first step of the identification
process. 
Chapman et al. (2005) got spectroscopic redshifts of 73 sources
obtained using the radio identification. The redshift distribution peaks 
at z= 2.2 with a substantial
tail up to z=4. Almost all SMGs are found in the redshift range $1.5<z<3$. 
\\

Many LIRGs and ULIRGs at low redshifts have been identified with
interacting or galaxy mergers. A substantial fraction show signs of
AGN activity but it has been shown for the low-redshift LIRGs and
ULIRGs that the starburst component dominates the energy output
(Genzel et al. 1998; Lutz et al. 1998). The sources used for the
redshift distribution of Chapman et al. (2005) have been imaged with
the HST.  Most of them are multi-component-distorted galaxy systems
(Conselice et al. 2003; Smail et al. 2004). They display irregular and
frequently highly complex morphologies compared to optically selected
galaxies at similar redshifts. They are often red galaxies with bluer
companions, as expected for interacting, star-forming galaxies.  They
have higher concentrations, and more prevalent major-merger
configurations than optically-selected galaxies at $z\sim$2-3.  Most
strikingly, most of the SMGs are extraordinarily large and elongated
relative to the field population regardless of optical magnitude.
SMGs have large bolometric luminosities, $\sim
10^{12}-10^{13}$ L$_{\odot}$, characteristic of ULIRGs. If the
far-infrared emission arises from the star formation, the large
luminosities translate to very high SFR $\ge$1000 M$_{\odot}$
year$^{-1}$. Such high rates are sufficient to form the stellar
population of a massive elliptical galaxy in only a few dynamical
times, given a sufficient gas reservoir.  SMGs are very massive
systems with typical mass of 1-2$\times$10$^{11}$L$_{\odot}$ (Swinbank
et al. 2004), comparable to the dynamical mass estimates from CO
observations.  Genzel et al. (2004; and more recently Greve et
al. 2005) have undertaken an ambitious program to study the nature of
the SMGs in more details. They got CO spectra with the Plateau de Bure
interferometer for 7 sources out of their sample of 12 for the CO 3-2
and 4-3 transitions redshifted in the 3~mm atmospheric window.  They
provide optical identifications and redshifts.  The detection of these
sources at the proper redshift confirms the usefulness of
identification with the help of the radio sources.  The median
redshift of this sample is 2.4.  In addition, one source was studied
with the SPIFI instrument on the ESO/VLT.  These observations are
giving very interesting clues on the nature of the submillimeter
galaxies.  The gas masses obtained for these systems using CO
luminosity/mass of gas determined from local ULIRGs is very large with
a median of 2.2$\times$$10^{10} M_{\odot}$ (10 times larger than in
the Milky Way).  Using the velocity dispersion, they could infer that
the dynamical median mass of these systems is 13 times larger than in
Lyman-break galaxies (LBGs) at the same redshift or 5 times the mass
of optically selected galaxies at this redshift.  These SMGs with a
flux at 850~$\mu$m larger than 5~mJy are not very rare and unusual
objects, because they contribute to about $20 \%$ of the CIB at this
frequency.  Through multiwavelength observations, Genzel et al. (2004)
get the stellar component in K band, and infer the star-formation rate
and duration of the star-formation burst. They can then compare the
number density of these massive systems with semiempirical models of
galaxy formation. The very interesting result is that this number
density is significantly larger than the predicted one, although the
absolute numbers depends on a number of assumptions like the IMF.
Such massive systems
at high redshift are not easy to understand in current cold dark
matter hierarchical merger cosmogonies. However, one must keep in mind
that bright SMGs (S$_{850}>$5~mJy) that contribute $20 \%$ of the CIB
may not be representative of the whole population. Gravitational lens
magnification provides a rare opportunity to probe the nature of the
distant sub-mJy SMGs. Kneib et al. (2005) study the property of one
SMG with an 850~$\mu$m flux S$_{850}$=0.8 mJy at a redshift of
$z=2.5$. This galaxy is much less luminous and massive than other
high-$z$ SMGs. It resembles to similarly luminous dusty starbursts
resulting from lower-mass mergers in the local Universe.\\

\begin{figure}[top]  
\vspace*{1.25cm}  
\begin{center}
\epsfig{figure=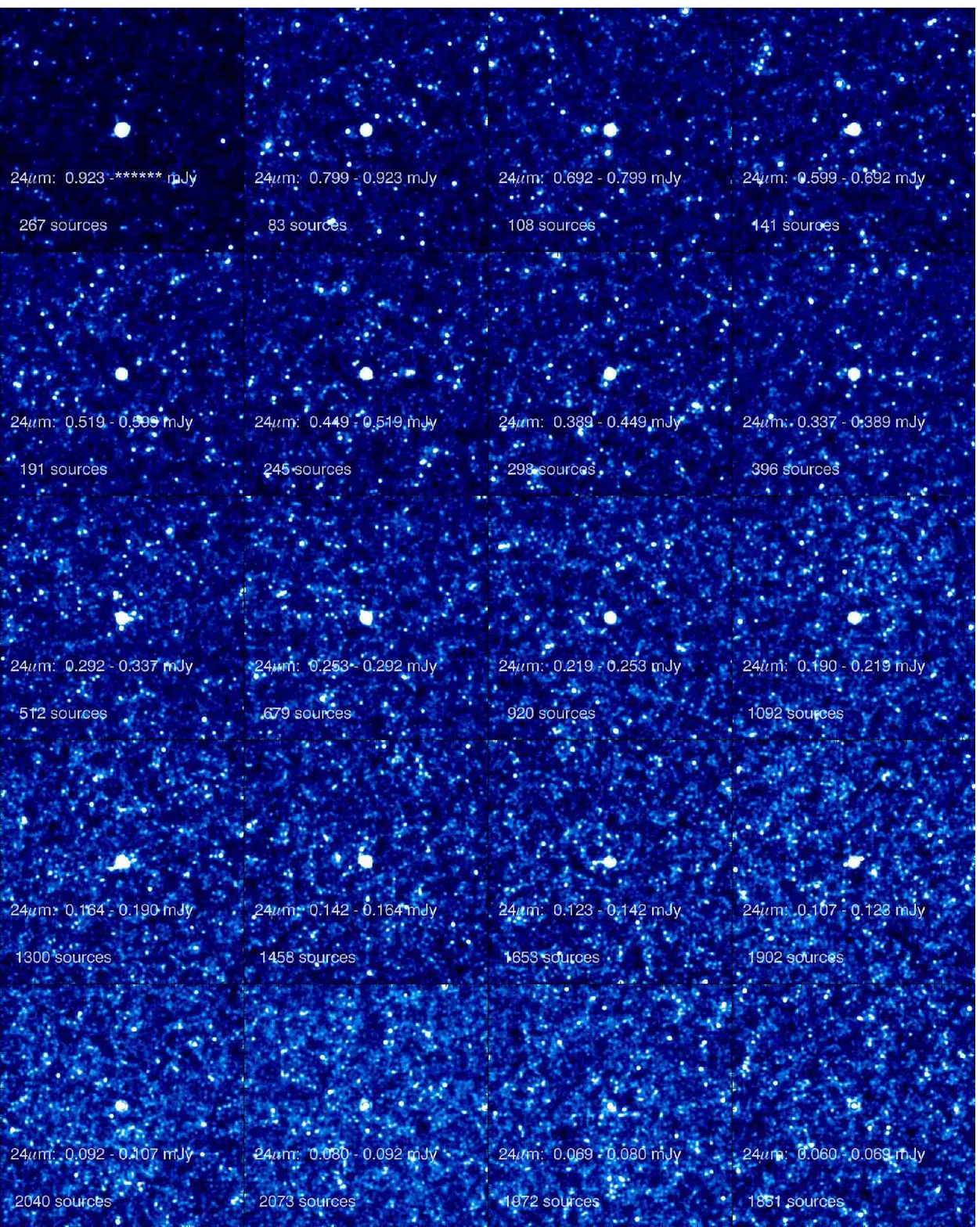,width=8.3cm}  
\epsfig{figure=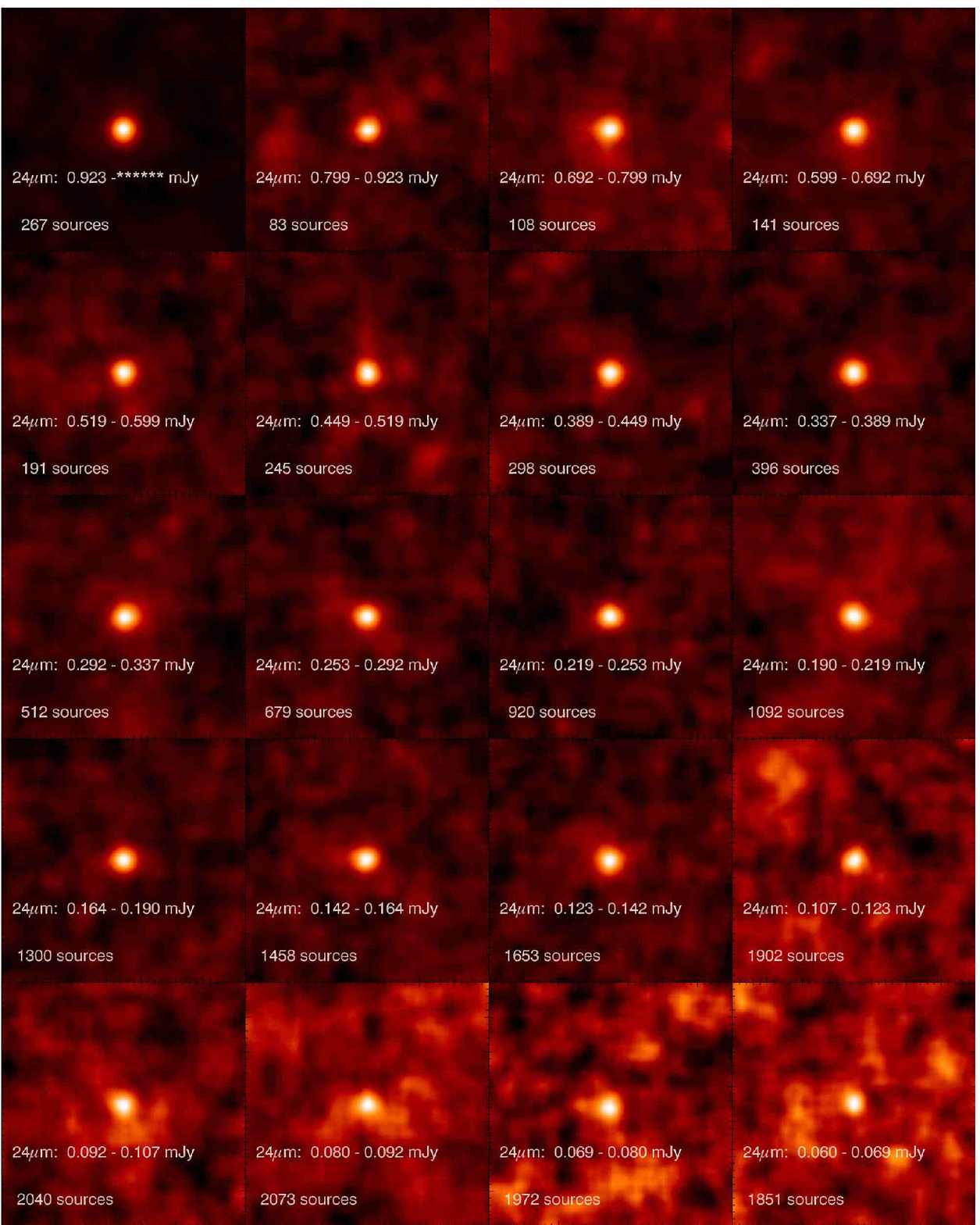,width=8.3cm}  
\end{center}
\caption{\label{fig_stacked} {\it Left:} Images of stacked {\it Spitzer} 24 $\mu$m sources per bin
of 24 $\mu$m flux. More than 19000 sources has been used. {\it Right:} The corresponding
stacked images at 160 $\mu$m.
} 
\end{figure} 

In order to link the different population of high-redshift objects,
several LBGs at redshift between 2.5 and 4.5 have been targeted at
850~$\mu$m. The Lyman-break technique (Steidel et al. 1996) detects
the rest-frame 91.2 nm neutral hydrogen absorption break in the SED of
a galaxy as it passes through several broad-band filters. LBGs are the
largest sample of spectroscopically confirmed high-redshift
galaxies. Observing LBGs in the submillimeter is an important goal,
because it would investigate the link, if any, between the two
populations. However, the rather low success rate of submillimeter
counterpart of LBGs (e.g., Chapman et al. 2000; Webb et al. 2003)
argues against a large overlap of the two populations.

\subsection{SPITZER observations: linking the ISOCAM galaxies and the SMGs}
A potential new way to find high-$z$ LIRGs and ULIRGs appeared
recently with the launch of the {\it Spitzer}
observatory. Particularly suited to this goal is the 24~$\mu$m channel
of the MIPS instrument.
Le Floc'h et al. (2004) give the first hint on the 24~$\mu$m selected
galaxies. They couple deep 24~$\mu$m observations in the Lockman hole
and extended groth strip with optical and near-infrared data to get
both identification and redshift (either spectroscopic or
photometric).  They find a clear class of galaxies with redshift 1$\le
z \le$2.5 and with luminosities greater than $\sim$5$\times$10$^{11}$
L$_{\odot}$ (see also Lonsdale et al. 2004). These galaxies are rather
red and massive with M$>$2$\times$10$^{10}$ M$_{\odot}$ (Caputi et
al. 2005). Massive star-forming galaxies revealed at $2\leq z \leq 3$
by the 24~$\mu$m deep surveys are characterized by very high star
formation rates -- SFR $\geq$500~M$_{\odot}$ year$^{-1}$.  They are
able to construct a mass of $\simeq$10$^{11}$~M$_{\odot}$ in a burst
lifetime ($\simeq$0.1 Gyr). The 24~$\mu$m galaxy population also
comprises sources with intermediate luminosities (10$^{10} \leq L_{IR}
\leq$10$^{11}$ L$_{\odot}$) and low to intermediate assembled stellar
masses (10$^{9} \leq$M$\leq$10$^{11}$ M$_{\odot}$) at $z\leq0.8$. At
low redshifts, however, massive galaxies are also present, but appear
to be building their stars quiescently in long timescales (Caputi et
al. 2005).  At these redshifts, the efficiency of the burst-like mode
is limited to low mass M$\leq$10$^{10}$ M$_{\odot}$ galaxies. These
results support a scenario where star-formation activity is
differential with assembled stellar mass and redshift, and proceed
very efficiently in massive galaxies (Caputi et al. 2005).
In the Lockman Hole, only one galaxy is associated with an X-ray
source. This suggests that these galaxies are mostly dominating by
star formation, consistent with the findings of Alonso-Herrero et
al.(2004) and Caputi et al. (2005). This is also suggested by SEDs
that are best fitted by PAH features rather than by strongly rising,
AGN-type continua (Elbaz et al. 2005).  The selected sources exhibit a
rather wide range of MIPS to IRAC flux ratio and optical/near-infrared
shapes, suggesting a possibly large diversity in the properties of
infrared galaxies at high redshift as noticed by Yan et
al. (2004). Based on these first analyzes, together with the
interpretation of the number counts (e.g., Lagache et al. 2004), it is
clear that the 24~$\mu$m observations will provide the sample to
unambiguously characterize the infrared galaxies up to
$z\simeq2.5$.

\begin{figure}[top]  
\vspace*{1.25cm}  
\begin{center}
\epsfig{figure=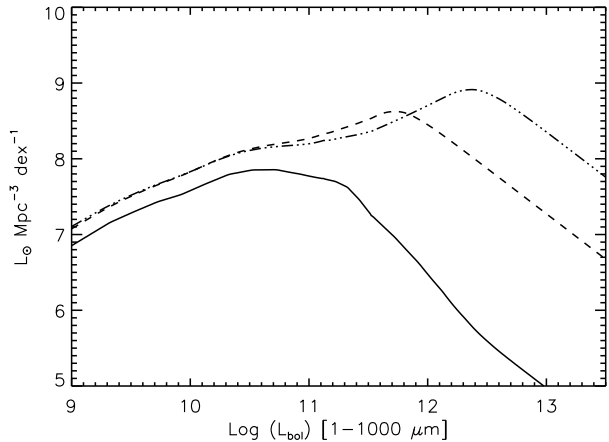,width=9cm}  
\end{center}
\caption{\label{fig:IRoutput} Co-moving evolution of the IR energy output per dex of
luminosity (L dN/dLogL). 
The continuous, dashed, dashed-3dotted 
lines are for z=0, 0.5 and 2 respectively. From Lagache et al. (2004).
} 
\end{figure} 

\section{Towards resolving the unresolved CIB}
As seen in Sect. \ref{Resolved_CIB}, despite the great sensitivity and the
high S/N ratio achieved by the recent {\it Spitzer} surveys at 70 and 160 $\mu$m,
the bulk of the CIB in energy, i.e., the peak at about 150 $\rm \mu$m, is not
resolved in individual sources (i.e. less than 10\% at 160 $\mu$m). This is due to the confusion.
The dominant contribution at the CIB peak can be inferred from the ISOCAM deep surveys
(Sect. \ref{zdistribCIB})
but this approach is model-dependent since it uses template SEDs to extrapolate the
fluxes measured at
15 $\mu$m to the peak of the CIB.
Dole et al. (2005) conduct an original approach. They use a stacking analysis method
that takes advantage of the good {\it Spitzer} 24 $\mu$m channel to fill the gap between the mid-IR
and the far-IR surveys. Down to a flux of 60 $\mu$Jy, {\it Spitzer} resolved 79\% of the CIB at
24 $\mu$m. The process of the stacking analysis is the following. First, the 24 $\mu$m sources
are sorted by decreasing fluxes and are put in 20 bins of flux with roughly equal
logarithmic width of $\Delta S_{24} / S_{24} \sim 0.15 $ (except for the bin corresponding
to the brightest flux). A square image, centered on each 24 $\mu$m source is then extracted
at both the 24, 70 and 160 $\mu$m. The images are finally added to generate a stacked
image of sources at the three wavelengths for each 24 $\mu$m flux bin (Figure \ref{fig_stacked}).
Surprisingly, stacked sources are detected at both 70 and 160 $\mu$m up to the faintest
24 $\mu$m flux bin. The fraction of resolved CIB (using the Lagache et al. (2004) CIB values)
becomes 83\% and 66\% at 70 and 160 $\mu$m, respectively. If new CIB estimates
from Dole et al. (2005) are used instead, the fraction of resolved CIB  becomes more than
75\% at both wavelengths. 
Dole et al. (2005) thus directly measure that a significant fraction of the far-IR CIB
is resolved using the mid-IR galaxies, without any model, nor hypothesis. Deriving the properties
of the 24 $\mu$m galaxies will thus directly probe the bulk of the CIB.\\

The stacking anlysis also allows for the first time to obtain average SEDs of galaxies from z=0 to z~2
by stacking the 24 $\mu$m galaxies per redshift bin in the Chandra Deep Field South, using the
redshifts derived by Caputi et al. (2005). The derived SEDs confirm the strong luminosity
evolution from z=0 to 2 (as shown in the next section). The observed colors can be qualitatively
explained by dusty IR luminous galaxies with an important dust component. Comparison with
models is provided in Dole et al. (2005).

\section{Cosmic evolution of IR galaxies}
Another remarkable property of the IR sources is their
extremely high rates of evolution with redshift exceeding those
measured for galaxies at other wavelengths and comparable to or larger
than the evolution rates observed for quasars. As an example, number counts at 15
$\rm \mu$m show a prominent bump peaking at about 0.4~mJy.  At the
peak of the bump, the counts are one order of magnitude above the
non-evolution models. In fact, data require a combination of a
(1+z)$^3$ luminosity evolution and (1+z)$^3$ density evolution for the
starburst component at redshift lower than 0.9 to fit the strong
evolution. Although it has not been possible with ISOCAM to probe in
detail the evolution of the infrared luminosity function, 
{\it Spitzer} data at 24~$\rm \mu$m give for the first time tight
constraints up to redshift 1.2 (Le Floc'h et al. 2005;
P\'erez-Gonz\'alez et al. 2005). A strong evolution is noticeable and
requires a shift of the characteristic luminosity L$^{\star}$ by a
factor (1+z)$^{4.0\pm0.5}$. Le Floc'h et al. (2005) find that the
LIRGs and ULIRGs become the dominant population contributing to the
comoving infrared energy density beyond $z\sim$0.5-0.6 and represent
70\% of the star-forming activity at $z\sim$1. The comoving luminosity
density produced by luminous infrared galaxies was more than 10 times
larger at $z\sim$1 than in the local Universe. For comparison, the
B-band luminosity density was only three times larger at $z=1$ than
today. Such a large number density of LIRGs in the distant Universe
could be caused by episodic and violent star-formation events,
superimposed on relatively small levels of star formation activity.
These events can be associated to major changes
in the galaxy morphologies. The rapid decline of the luminosity
density from $z=1$ is only partially due to the decreasing frequency
of major merger events. Bell et al. (2005) showed that the SFR density
at $z\sim$0.7 is dominated by morphologically normal spiral, E/S0 and
irregular galaxies ($\ge$70\%), while clearly interacting galaxies
account for $<$30\%. The dominent driver of the decline is a strong
decrease in SFR in morphologically undisturbed galaxies.  This could
be due, for example, to gas consumption or to the decrease of weak
interactions with small satellites that could trigger the star
formation through bars and spiral arms.\\

At still higher redshifts, the infrared
luminosity of the sources that dominate the background is larger than
$10^{12} L_{\odot}$. This is a population with a very different
infrared luminosity function than the local or even the $z=1$
luminosity function. The global evolution of the IR energy output with redshift
is illustrated on Figure \ref{fig:IRoutput}.

\section{Conclusions and Challenges}
A number of conclusions are now clear from the analysis of the identified 
sources in the CIB:
\par\medskip\noindent
$\bullet$ The comoving energy produced in the past that makes up the CIB at different wavelengths 
is more uniform that what is suggested by its spectral energy
distribution. This is due to the fact that the CIB at long wavelengths
($\lambda \ge 400 \mu$m) is dominated by emission from the peak of the
SED of galaxies at high $z$. More quantitatively, the ISOCAM surveys
reveal that about two-thirds of the CIB emission at $\lambda \sim$150
$\mu$m is generated by LIRGs at $z \sim 0.7$. At 850~$\mu$m, more than
half of the submillimeter CIB is generated by SMGs. The brightest SMGs
($S_{850}>$3 mJy, $\sim$30\% of the CIB) are ULIRGs at a median
redshift of 2.2. The energy density at 150~$\mu$m, which is
$\sim$20-25 times larger than the energy density at 850~$\mu$m
requires a comoving energy production rate at $z=0.7$ roughly 10 times
the energy production rate at $z=2.2$.
\par\medskip\noindent
$\bullet$ The evolution exhibited by LIRGs and ULIRGs is much faster than for optically 
selected galaxies. The ratio of infrared to optical, volume-averaged
output of galaxies increases rapidly with increasing redshift.
\par\medskip\noindent
$\bullet$ Luminosity function evolution is such that the power output is dominated by
LIRGs at $z\simeq 0.7$ and ULIRGs at $z\simeq 2.5$.
\par\medskip\noindent
$\bullet$  The energy output of CIB sources is dominated by starburst activity.
\par\medskip\noindent
$\bullet$ AGN activity is very common in the most luminous of these galaxies even though this 
activity does not dominate the energy output. The rate and fraction of
the energy produced increase with the luminosity.
\par\medskip\noindent
$\bullet$ LIRGs at $z\simeq 0.7$ are dominated by interacting massive
late-type galaxies. They seem to be starburst phases of
already-built massive, late-type field galaxies accreting gas or
gas-rich companions forming the disks. We see today a rapid decrease
of this activity probably associated with a dry out of the gas
reservoir in their vicinity.  The larger redshift IR galaxies ($z\simeq 2.5$), 
which are also
more luminous, seem to belong to more massive complex systems
involving major merging. These systems could be located in the rare
larger amplitude peaks of the large-scale structures leading to
massive elliptical galaxies at the center of rich clusters.
\par\medskip\noindent
$\bullet$  SMGs show rather strong correlations with correlation lengths larger than those of other
high redshift sources.
\par\medskip\noindent
$\bullet$ LIRGs and ULIRGs cannot be identified with any of the distant 
populations found by rest-frame ultraviolet and optical surveys.\\

Although these findings are answering the basic questions about the
sources that make up the CIB, there are still observational
difficulties to be overcome to complete these answers.  The SEDs of
LIRGs and ULIRGs are quite variable and often not very well
constrained in their ratio of far-infrared to mid-infrared or to
submillimeter wavelengths.  The far-infrared, where most of the energy
is radiated, requires cryogenically cooled telescopes. These have
small diameters and, hence, poor angular resolution and severe
confusion limits for blind surveys. Establishing proper SEDs for the
different classes of infrared galaxies detected either in mid-infrared
(with ISOCAM at 15~$\mu$m or MIPS at 24~$\mu$m) or in
millimeter-submillimeter surveys is one of the challenges of the
coming decade.  Making sure that no class of sources that contribute
significantly to the CIB at any wavelength has been missed is an other
observational challenge. The submillimeter galaxies not found through
the radio-selected sources and the question of the warm submillimeter
galaxies are also two of those challenges.


\vfill 
\end{document}